\documentclass[aps,prd,preprint,superscriptaddress,showpacs]{revtex4}
\usepackage{epsfig}
\usepackage{graphicx}
\usepackage{ulem}
\usepackage{color}
\definecolor{My_red}{cmyk}{0.00,1.00,1.00,0.20}

\def\bwt{\begin{widetext}}
\def\ewt{\end{widetext}}
\def\be{\begin{equation}}
\def\ee{\end{equation}}
\def\bea{\begin{eqnarray}}
\def\eea{\end{eqnarray}}
\def\bean{\begin{eqnarray*}}
\def\eean{\end{eqnarray*}}
\def\bary{\begin{array}}
\def\eary{\end{array}}
\def\bit{\begin{itemize}}
\def\eit{\end{itemize}}

\def\nn{\nonumber}

\def\su5u1{SU(5) \times U(1)}
\def\fsu5u1{SU(5) \times U(1)'}
\def\fsu5{$\cal F$-$SU(5)$}
\def\sq20{SO(10) \times SO(10)}

\begin{document}
\title{Rare B decays  in the \fsu5 Model  }

\affiliation{Institute of Theoretical Physics,
     Beijing University of Technology, Beijing 100124, China}

\author{Tianjun Li}

\affiliation{ State Key Laboratory of Theoretical Physics,
      Institute of Theoretical Physics, Chinese Academy of Sciences,
Beijing 100190, P. R. China }

\affiliation{George P. and Cynthia W. Mitchell Institute for
Fundamental Physics and Astronomy, Texas A$\&$M University,
College Station, TX 77843, USA}

\author{Dimitri V. Nanopoulos}

\affiliation{George P. and Cynthia W. Mitchell Institute for
Fundamental Physics and Astronomy, Texas A$\&$M University,
College Station, TX 77843, USA}

\affiliation{Astroparticle Physics Group,
Houston Advanced Research Center (HARC),
Mitchell Campus, Woodlands, TX 77381, USA}

\affiliation{Academy of Athens, Division of Natural Sciences, 28
Panepistimiou Avenue, Athens 10679, Greece}

\author{Wenyu Wang }
\affiliation{Institute of Theoretical Physics,
     Beijing University of Technology, Beijing 100124, China}

\author{ Xiao-Chuan Wang }
\affiliation{Institute of Theoretical Physics,
     Beijing University of Technology, Beijing 100124, China}

\author{ Zhao-Hua Xiong}
\affiliation{Institute of Theoretical Physics,
     Beijing University of Technology, Beijing 100124, China}

\date{\today}
\begin{abstract}
In the testable Flipped $SU(5)\times U(1)_X$ model with TeV-scale vector-like particles from F-theory
model building dubbed as the \fsu5 model, we study the vector-like quark contributions to B
physics processes, including the quark mass spectra, Feynman rules, new operators and Wilson
coefficients, etc. We focus on the implications
of the  vector-like quark mass scale  on
B physics. We find that there exists the $\bar{s}bZ$ interaction
at tree level, and the Yukawa interactions are changed.  Interestingly,
different from many previous models, the effects of vector-like quarks
on rare B decays such as $B\to X_s\gamma$ and $B\to X_s\ell^+\ell^-$
do not decouple in some viable parameter space, especially
when the vector-like quark masses are comparable to the charged Higgs boson mass.
Under the constraints  from $B\to X_s\gamma$ and $\ B\to X_s\ell^+\ell^-$,
the latest measurement for  $B_s\to \mu^+\mu^-$ can be explained naturally,
 and the branching ratio of $B_s\to \ell^+\ell^-\gamma$ can be up to
$(4\sim5)\times10^{-8}$. The non-decouling effects are much
more predictable and thus the \fsu5 model may be tested in the
near future experiments.

\end{abstract}

\preprint{ACT-05-12, MIFPA-12-16}

\maketitle

\section{Introduction}

Supersymmetry provides a natural solution to the
gauge hierarchy problem in the Standard Model (SM).
In the supersymmetric SM (SSM) with R-parity
under which the SM particles are even
while the supersymmetric particles (sparticles)
are odd, the $SU(3)_C\times SU(2)_L\times U(1)_Y$ gauge
couplings can be unified around $2\times 10^{16}$
GeV~\cite{Langacker:1991an}, the lightest supersymmetric
particle (LSP) such as the
 neutralino can be a cold dark matter
candidate~\cite{Ellis:1983ew, Goldberg:1983nd},
and the electroweak (EW) precision constraints can be
evaded, etc. Especially, the gauge coupling unification strongly
suggests Grand Unified Theories (GUTs).
However, in the supersymmetric $SU(5)$ models,
there exist the doublet-triplet splitting problem and
dimension-five proton decay problem. Interestingly, these problems
can be solved elegantly in the Flipped $SU(5)\times U(1)_X$
models~\cite{smbarr, dimitri, AEHN-0} via missing partner mechanism~\cite{AEHN-0}.
On the other hand, string theory is the most promising candidate
for quantum gravity, and it can unify all the fundamental interactions
in the Nature. However, the string scale is  at least one-order larger
than the conventional GUT scale.

To solve the little hierarchy problem between the traditional GUT scale
and string scale, two of us (TL and DVN) with Jing Jiang have proposed the
testable Flipped $SU(5)\times U(1)_X$ models, where the
TeV-scale vector-like particles are introduced~\cite{Jiang:2006hf}.
Such kind of models can be constructed
from the free fermionic string
constructions  at the Kac-Moody level one~\cite{Antoniadis:1988tt, Lopez:1992kg}
and locally  from
the F-theory model building~\cite{Beasley:2008dc, Jiang:2009zza},
and is dubbed as ${\cal F}$-$SU(5)$~\cite{Jiang:2009zza}.
In particular, these models are very
interesting from the phenomenological point of view~\cite{Jiang:2009zza}:
the vector-like particles can be observed at the Large Hadron Collider (LHC),
 proton decay is within the reach of the future
Hyper-Kamiokande~\cite{Nakamura:2003hk} and
Deep Underground Science and Engineering
Laboratory (DUSEL)~\cite{DUSEL} experiments~\cite{Li:2009fq, Li:2010dp},
the hybrid inflation can be naturally realized,  the
correct cosmic primodial density fluctuations can be
generated~\cite{Kyae:2005nv}, and the lightest CP-even Higgs boson
mass can be lifted~\cite{Huo:2011zt, Li:2011ab}.  With no-scale boundary conditions
at $SU(5)\times U(1)_X$ unification scale~\cite{Cremmer:1983bf},
two of us (TL and DVN) with James Maxin and Joel Walker have
 described an extraordinarily constrained ``golden point''~\cite{Li:2010ws}
and ``golden strip''~\cite{Li:2010mi} that satisfied all the latest
experimental constraints and has an imminently observable proton
decay rate~\cite{Li:2009fq}. For a review of the recent progresses,
see Ref.~\cite{Li:2012uj}.

Interestingly, the vector-like quarks in the \fsu5 model predict
 rich phenomenology on low energy processes.
 If the model is treated seriously,
 constraints from  electroweak parameters such as  $U,\ S, T$ and $R_b,\ R_c$
 and B processes should be taken into account.   We also would like to
 point out that the \fsu5 model has no Landau pole problem and then
is very different from the other simple SM extensions in quark sector (also see the next Section)~\cite{BNPfourth},
and  the $3\times 3$ SM-like quark mixing matrix
is now replaced by a $5\times 5$ one and then is no longer unitary, and
there exists the tree-level $\bar{s}bZ$ interaction, which will play an important,
even dominant, role in some parameter space for rare B decays.

Thanks to the efforts of the B factories and LHC,
the exploration of quark-flavor mixing is now entering
a new interesting era. It is well known that
the rare B decays induced by the flavor changing neutral current (FCNC)
only occur at loop level in the SM and then are sensitive to new physics. Thus, the
rare radiative, leptonic and semi-leptonic B meson decays are valuable in testing the SM at
loop level and probe new physics.
 On the theoretical side, the rare B inclusive radiative decays
 $B \to X_s \gamma$ and $B\to X_s\ell^+\ell^-(\ell=e,\mu)$ as well as
the exclusive decays $B_s\to \mu^+\mu^-$ and $B_s\to \ell^+\ell^-\gamma$ have been studied
extensively at the leading logarithm order (LO)~\cite{BLOSM} and high order
 in the SM~\cite{BHOSM} and various new physics models~\cite{BNPfourth, BNPMSSM, BNP2HDM}.
 On the experimental side, $B \to X_s \gamma$ and $B\to X_s \ell^+\ell^-(\ell=e,\mu)$   have been measured
 and the latest upper bound on $B_s\to \mu^+\mu^-$ is achieved~\cite{Bmeasured}.
 By comparing the predictions with experimental measurements,
 we will present some constraints on the parameter space in the \fsu5 model.

The first task of this work will be deriving the  quark mass spectra
and  Feynman rules. We stress that  the Feynman rules which not be
presented in previous studies are used not only in B physics but also
in research  of all low energy processes. B physics constraints on the model
is the second task of this work, we will concentrate our attention
on the vector-like quark contributions to B physics, in particular,
 the contributions from the new operators induced by tree-level FCNC.
We will show that the $\bar{s}bZ$ interaction can be generated
at tree level, and the Yukawa interactions are changed,  new operators
 $O_9'$ and $O_{10}'$ in effective Hamiltonian should  be introduced.
 We will demonstrate that different from many previous models, the effects of vector-like quarks
on rare B decays such as $B\to X_s\gamma$ and $B\to X_s\ell^+\ell^-$ do not decouple
 in some allowed parameter space, especially
when the vector-like quark masses are comparable to the charged Higgs boson mass.
Within the constraints  from $B\to X_s\gamma$ and $\ B\to X_s\ell^+\ell^-$, and
 the latest measurement for  $B_s\to \mu^+\mu^-$ will be explained naturally,
 and the branching ratio of $B_s\to \ell^+\ell^-\gamma$ can be up to
$(4\sim5)\times10^{-8}$. Because the non-decouling effects are very
 predictable, the \fsu5 model may be tested in the
near future experiments.

This paper is organized as follows.
We present a brief description for the TeV-scale \fsu5 model and derive
all the Feynman rules for our calculations in Section~\ref{model}. We
  discuss the implications of vector-like quarks on B physics in Section~\ref{bphys}.
Our numerical results are presented in Section~\ref{num},
and Section~\ref{sum} is the summary.

\section{The \fsu5 Model around the TeV Scale}\label{model}

To achieve the string-scale gauge coupling unification in the \fsu5 model, we introduce the
vector-like particles which from complete Flipped $SU(5)\times U(1)_X$ multiplets.
The quantum numbers for these additional vector-like particles
 under the $SU(5)\times U(1)_X$ gauge symmetry are~\cite{Jiang:2006hf}
\begin{eqnarray}
&& XF ={\mathbf{(10, 1)}}~,~{\overline{YF}}={\mathbf{({\overline{10}}, -1)}}~,~\nn\\
&& Xf={\mathbf{(5, 3)}}~,~{\overline{Yf}}={\mathbf{({\overline{5}}, -3)}}~,~\nn\\
&& Xl={\mathbf{(1, -5)}}~,~{\overline{Yl}}={\mathbf{(1, 5)}}~.~
\end{eqnarray}
To avoid the confusion in the following discussions,
we change the convention in Ref.~\cite{Jiang:2006hf} a little bit.
It is obvious that $XF$, ${\overline{YF}}$, $Xf$, ${\overline{Yf}}$,
$Xl$, and ${\overline{Yl}}$  are standard vector-like
particles with  contents as follows
\begin{eqnarray}
&& XF = (XQ, XD^c, XN^c)~,~ {\overline{YF}}=(YQ^c, YD, YN)~,~\nn\\
&& Xf=(XU, XL^c)~,~ {\overline{Yf}}= (YU^c, YL)~,~\nn\\
&& Xl= XE~,~ {\overline{Yl}}= YE^c~.~
\end{eqnarray}
Under the $SU(3)_C \times SU(2)_L \times U(1)_Y$ gauge
symmetry, the quantum numbers for the extra vector-like
particles are
\begin{eqnarray}
&& XQ={\mathbf{(3, 2, {1\over 6})}}~,~
YQ^c={\mathbf{({\bar 3}, 2,-{1\over 6})}} ~,~\nn\\
&& XU={\mathbf{({3},1, {2\over 3})}}~,~
YU^c={\mathbf{({\bar 3},  1, -{2\over 3})}}~,~\nn\\
&& XD={\mathbf{({3},1, -{1\over 3})}}~,~
YD^c={\mathbf{({\bar 3},  1, {1\over 3})}}~,~\nn\\
&& XL={\mathbf{({1},  2,-{1\over 2})}}~,~
YL^c={\mathbf{(1,  2, {1\over 2})}}~,~\nn\\
&& XE={\mathbf{({1},  1, {-1})}}~,~
YE^c={\mathbf{({1},  1, {1})}}~,~\nn\\
&& XN={\mathbf{({1},  1, {0})}}~,~
YN^c={\mathbf{({1},  1, {0})}}.
\end{eqnarray}
At the GUT scale the superpotential is given by
\begin{eqnarray}
W_{GUT}&=& Y_{ij}^{D}F_{i}F_{j}h+Y_{ij}^{U\nu}F_{i}\bar{f_{j}}\bar{h}
+Y_{ij}^{E}\bar{l_{i}}\bar{f_{j}}h
+\mu h\bar{h}+Y_{kj}^{N}\phi_{k}\bar{H}F_{j}\nonumber\\
&+&Y'{}_{j}^{D}XFF_{j}h+Y'{}_{j}^{U\nu}XF\bar{f}_{j}\bar{h}+Y"{}_{i}^{U\nu}F_{i}\overline{Xf}\bar{h}
+Y'{}_{j}^{E}\overline{Xl}\bar{f}_{j}h \nonumber\\
&+&Y"{}_{j}^{E}\bar{l}_{j}\overline{Xf}h+Y'{}_{k}^{N}\phi_{k}\bar{H}XF
+ Y^{2D}XFXFh+Y'^{2D}\overline{YF}\overline{YF}\bar{h}\nonumber\\
&+&Y^{2U\nu}XF\overline{Xf}\bar{h}+Y'^{2U\nu}\overline{YF}Yfh+Y{}^{2E}\overline{Xl}\overline{Xf}h
+Y'^{2E}YlYf\bar{h}\nonumber\\
&+& M_{j}^{1}F_{j}\overline{YF}+M_{j}^{2}\bar{f}_{j}Yf+M_{j}^{3}\bar{l}_{j}Yl
\nonumber\\
&+& M^{4}XF\overline{YF}+M^{5}\overline{Xf}Yf+M^{6}\overline{Xl}Yl~,
\end{eqnarray}
where $i$ is the generation indices. The first line  is the SSM superpotential,
the second line is the Yukawa mixing terms between the SM fermions and vector-like particles,
the third and fourth lines are the SM-like superpotential for vector-like multiplets,
and the fifth and sixth lines are bilinear mass terms.
After the $SU(5)\times U(1)_X$ gauge symmetry
breaking down to the SM gauge symmetry, we obtain the superpotential as follows
\begin{eqnarray}
W_{EW} &=& (Y_{ij}^{D}-Y_{ji}^{D})(D^{c})_{i}Q_{j}\cdot H_{d}+Y_{ij}^{U\nu}U_{j}^{c}Q_{i}\cdot H_{u}
-Y_{ij}^{U\nu}N_{i}^{c}L_{j}\cdot H_{u}\nonumber\\
&-&Y_{ij}^{E}E_{i}^{c}L\cdot H_{d}-Y_{j}^{'D}(XD^{c}Q_{j}\cdot H_{d}+D_{j}^{c}XQ\cdot H_{d})
+Y_{'j}^{U\nu}U_{j}^{c}XQ\cdot H_{u}\nonumber\\
&-&Y_{j}^{'U\nu}XN^{c}L\cdot H_{u}+Y_{i}^{"U\nu}XU^{c}Q\cdot H_{u}
-Y_{i}^{"U\nu}N_{i}^{c}XL\cdot H_{u}-Y_{j}^{'E}XE^{c}L\cdot H_{d}\nonumber\\
&-&Y_{j}^{"E}E_{j}^{c}XL\cdot H_{d}-2Y^{2D}XD^{c}XQ\cdot H_{d}-2Y^{'2D}YDYQ^{c}\cdot H_{u}\nonumber\\
&+&Y^{2U\nu}XU^{c}XQ\cdot H_{u}-Y^{2U\nu}XN^{c}XL\cdot H_{u}-Y^{'2U\nu}YUYQ^{c}\cdot H_{d}\nonumber\\
&+&Y^{'2U\nu}YNYL^{c}\cdot H_{d}-Y^{2E}XE^{c}XL\cdot H_{d}
-Y^{'2E}YEYL^{c}\cdot H_{u}\nonumber\\
&-&2M_{j}^{1}\left[D_{j}^{c}YD+Q\cdot YQ^{c}+N_{j}^{c}YN\right]+M_{j}^{2}\left[U^{c}YU+L\cdot YL^{c}\right]
+M_{j}^{3}E_{j}^{c}YE\nonumber\\
&-&2M^{4}\left[XD^{c}YD+XQ\cdot YQ^{c}+XN^{c}YN\right]+M^{5}\left[XU^{c}YU
+ XL\cdot YL^{c}\right]
\nonumber\\
&+& M^{6}XE^{c}YE~.~\,
\end{eqnarray}

At low energy, the sparticles decouple rapidly
when $M_S$ increases. Note that the LHC already put strong
constraints on squark masses around 1500~GeV,
 we will concentrate on the contributions from new
vector-like quark multiplets $XU,~YU^c,~XD,~{\rm and}~YD^c$ for simplicity.
At first glance these multiplets seem to be similar to
the fourth and fifth generation quarks, but indeed
$(XU,~YU^c)$ and $(XD,~YD^c)$ are vector-like.
This makes them very different from the fourth and fifth generation quarks.
The  down-type quark mass matrix is
\begin{eqnarray}
 M_D=&\left(\begin{array}{ccccc}
(Y_{11}^{D}+Y_{11}^{D})v_{d} & (Y_{12}^{D}+Y_{21}^{D})v_{d} &
(Y_{13}^{D}+Y_{31}^{D})v_{d} & Y'{}_{1}^{D}v_{d} & -2M_{1}^{1}\\
(Y_{21}^{D}+Y_{12}^{D})v_{d} & (Y_{22}^{D}+Y_{22}^{D})v_{d} &
(Y_{23}^{D}+Y_{32}^{D})v_{d} & Y'{}_{2}^{D}v_{d} & -2M_{2}^{1}\\
(Y_{31}^{D}+Y_{13}^{D})v_{d} & (Y_{32}^{D}+Y_{23}^{D})v_{d} &
(Y_{33}^{D}+Y_{33}^{D})v_{d} & Y'{}_{3}^{D}v_{d} & -2M_{3}^{1}\\
Y'{}_{1}^{D}v_{d} & Y'{}_{2}^{D}v_{d} & Y'{}_{3}^{D}v_{d} & 2Y^{2D}v_{d} & -2M^{4}\\
2M_{1}^{1} & 2M_{2}^{1} & 2M_{3}^{1} & 2M^{4} & -2Y'^{2D}v_{u}\end{array}\right)
\label{downmatrix}~,~
\end{eqnarray}
 and the up-type quark matrix is
 \begin{eqnarray}
  M_U= \left(\begin{array}{ccccc}
  Y_{11}^{U\nu}v_{u} & Y_{21}^{U\nu}v_{u} & Y_{31}^{U\nu}v_{u} & Y'{}_{1}^{U\nu}v_{u} & M_{1}^{2}\\
  Y_{12}^{U\nu}v_{u} & Y_{22}^{U\nu}v_{u} & Y_{32}^{U\nu}v_{u} & Y'{}_{2}^{U\nu}v_{u} & M_{2}^{2}\\
  Y_{13}^{U\nu}v_{u} & Y_{23}^{U\nu}v_{u} & Y_{33}^{U\nu}v_{u} & Y'{}_{3}^{U\nu}v_{u} & M_{3}^{2}\\
  Y"{}_{1}^{U\nu}v_{u} & Y"{}_{2}^{U\nu}v_{u} & Y"{}_{3}^{U\nu}v_{u} & Y^{2U\nu}v_{u} & M^{5}\\
  -2M_{1}^{1} & -2M_{2}^{1} & -2M_{3}^{1} & -2M^{4} & Y'^{2U\nu}v_{d}\end{array}\right)~,~
\label{upmatrix}
 \end{eqnarray}
 where $v_u$ and $v_d$ are the vacuum expectation values (VEVs) for $H_u$ and $H_d$.
These two matrixes can be diagonalized by unitary matrices $U$ and $V$,
\begin{eqnarray}
V_d^\dagger M_DU_d={\rm diag.}[m_d,m_s,m_b,m_{d_x},m_{d_y}],\nonumber\\
V_u^\dagger M_UU_u={\rm diag.}[m_u,m_c,m_t,m_{u_x},m_{u_y}].
\label{massdiag}
 \end{eqnarray}
Thus, the quark mixings are described by a matrix $V=U_u^\dagger U_d$.
From Eqs. (\ref{downmatrix}) and (\ref{upmatrix}),
we can see that the mass matrices of the down-type quarks and up-type quarks
are related to each other, implying that the Yukawa couplings are
different from those in the SM.  In the Feynman gauge the Feynman
rules for charged $W$ boson, Goldstone boson, and charged Higgs boson with quarks
$\overline{u_l}d_j \chi^+ (\chi=W,~G,~h)$
and for $Z$ boson $\overline{d_j}d_l Z$
needed in our calculations are given as follows
\begin{eqnarray}
&& i\frac{g}{\sqrt{2}}\gamma^\mu \left[g^\chi_L(l,j)P_L
+g^\chi_R(l,j)P_R\right],~~(\chi=W,~Z)~,\label{gud}\\
&& i\frac{g}{\sqrt{2}}\left[g^\chi_L(l,j)P_L
+g^\chi_R(l,j)P_R\right],~~(\chi=G,~h)~,
\label{hud}
\end{eqnarray}
where
\begin{eqnarray}
 g^W_L(i,j) &=& \sum_{m=1}^4U_u^{*mi} U_d^{m,j},\ \
  \ g^W_R(i,j) =V_u^{*5 i}V_d^{5j}, \label{ud_1}\\
 g^Z_L(i,j) &=& -\frac{1}{\sqrt{2}\cos\theta_{W}}\left[\left(1-\frac{2}{3}\sin^{2}
 \theta_{W}\right)\delta^{ij}-U_{d}^{*5i}U_{d}^{5j}\right],\nn\\
 g^Z_R(i,j) &=&  -\frac{1}{\sqrt{2}\cos\theta_{W}}\left[-\frac{2}{3}\sin^{2}
 \theta_{W}\delta^{ij}+V_{d}^{*5i}V_{d}^{5j}\right] , \label{ud_2}\\
g^G_L(i,j) &=& \left(\sum_{k,m=1}^4
Y^{U\nu}_{km}V_u^{*ki} U_d^{mj} + 2Y'^{2D}V_u^{*5i} U_d^{5j}\right)\frac{v_u}{m_W},\nn\\
g^G_R(i,j) &=& -\left(\sum_{k,m=1}^4
(Y^{D}_{mk}+Y^{D}_{km})V_d^{*k j} U_u^{mi} - 2Y'^{U\nu}V_d^{*5j} U_d^{5i}\right)\frac{v_d}{m_W},
 \label{ud_3}\\
g^{h}_L(i,j) &= & \left(\sum_{k,m=1}^4
Y^{U\nu}_{km}V_u^{*k i} U_d^{mj} + 2Y'^{2D}V_u^{*5 i} U_d^{5j}\right)\frac{v_d}{m_W},\nn\\
g^{h}_R(i,j) &=& \left(\sum_{k,m=1}^4
(Y^{D}_{mk}+Y^{D}_{km})V_d^{*kj} U_u^{mi} - 2Y'^{U\nu}V_d^{*5j} U_d^{5i}\right)\frac{v_u}{m_W}.
 \label{ud_4}
\end{eqnarray}
Because the vector-like particles do not change $U(1)_{EM}$ interaction,
the interactions of photon and quarks are still the same as those in the SM. From the above mass
matrices  we can see that the TeV-scale \fsu5 model has two points for rich physics to be explored:
\begin{itemize}
\item Since the quark mass matrices  are not the same as two Higgs doublet
model (2HDM)~\cite{BNP2HDM} or the Minimal Supersymmetric Standard Model (MSSM)~\cite{BNPMSSM},
the loop-level FCNC will be changed by the Yukawa interactions, and then  may change the prediction of
process $b \to s \gamma$ significantly.
\item The  last terms in Eqs.(\ref{ud_1})-(\ref{ud_4}), which we call the ``tail terms'',
 will cause the tree-level
FCNC processes induced by $b \to s\ell^+\ell^-$  and then the
stringent constraints on the model parameter space will be expected.
\end{itemize}

\section{Implications on B physics}\label{bphys}

Apart from the directly search for the light vector-like quarks at the LHC,
another way to  test the \fsu5 model is to measure their effects on low energy
processes  such as rare B decays.

\subsection{ Effective Hamiltonian}

The starting point for rare B decays  $B\to X_s \gamma$, $B\to X_s\ell^+\ell^-$, $B_s\to\ell^+\ell^-$
and  $B_s\to \ell^+\ell^-\gamma$  is the determination of a low-energy effective Hamiltonian obtained by
integrating out the heavy degrees of freedom in the theory.
For $b \to s $ transition, this can be written as
\begin{equation}
{\cal H}_{\rm eff} = - \frac{G_F}{\sqrt{2}} V_{ts}^* V_{tb}
\sum_{i=1}^{10} [C_i(\mu) O_i(\mu)+C^{'}_i(\mu) O^{'}_i(\mu)]~,~\,
\label{eq:HeffBXsgamma}
\end{equation}
where the effective operators $O_i$  are same as those in the SM defined in Ref.~\cite{BLOSM}.
The chirality-flipped operators $O'_i$ are obtained from $O_i$
by the replacement $\gamma_5\to -\gamma_5$ in quark current.
It is obvious that $O'_{9,10}$ can be got directly from the tail
terms in the Feynman rules of the \fsu5 model.
A few remarks follow on the operators and Wilson coefficients:
\begin{itemize}
\item  As mentioned in introduction, the three generation quark mixing
       matrix is replaced by a $5\times 5$ matrix $U_u^\dagger U_d$
       and then is non-unitary.
       In our analyses we take a reasonable assumption that the deviation
       from unitary is not large. Otherwise, the tree-level FCNC will modify
       significantly the low energy processes such
       as $Z\to b \overline {b}$ and $B_s\to \mu^+\mu^-$.
\item  Since the Wilson coefficient $C_2(m_W)=-\frac{V_{cb}V_{cs}^*}{V_{tb}V_{ts}^*}\simeq 1$
       is always a good approximation in
       \fsu5 model, and the coefficients of four quark operators $C_i(\mu_b)\ (i=1,3-6)$
       depend actually on the value $C_2(m_W)$, the contributions from
       the  four-quark operator matrix elements
       to effective coefficient $C_9^{eff}(\mu_b)$ can not be ignored
       and have the same expressions as the SM.
\item The coefficient of operator $O_2' =  (\overline{s} c)_{V+A}  (\overline{c} b)_{V-A}$,
      for example, is proportional to
      the elements of quark mixing matrix $V^{5j}_u$ or $U^{5i}_d$ denoted the mixings between
      the ordinary quarks and vector-like quarks. Thus, it  can be reasonably
      set to be  much smaller than $\mathcal{O}(1)$, and
      the contributions from
      the four-quark primed operators to $C_9^{eff}(\mu_b)$
      and $C_9^{',eff}(\mu_b)$ can be neglected safely.  This means
      \begin{equation}
      C_{9,10}^{',eff}(\mu_b)=C_{9,10}^{'}(m_W)~,~\,
      \end{equation}
      which receive contributions mainly from the tree-level diagrams,
      loop diagrams for $b\to s\gamma$, and box diagrams. We also neglect the
      operator $O_7'$ contribution.
\item For $b \to s \gamma$,  the new contributions mainly come
 from the new type Yukawa interactions, and for $b \to s \ell^+\ell^-$,
 the new contributions mainly arise from the new operators $O_{9,10}'$.
\end{itemize}

\subsection{Analyses in B Physics Calculations}
In the \fsu5 model the contributions to operators $O_i\ (i=1-10)$ and $O'_{9,10}$
can be encoded  by the values of the coefficients $C_{i}$
and $C'_i$ at the matching  scale $m_W$. In this Section, we will present
the Wilson coefficients at the matching scale and decay widths for some rare B decays.
 We keep both new physics contributions and the SM results at the LO for consistency.
\begin{itemize}
\item The Wilson coefficient $C_7$ at the matching scale is
  \begin{eqnarray}
  C_{7}&=&\frac{1}{V_{tb}V_{ts}^{\ast}}\sum_{i=1}^{5}\{ A(x_i)g_{L}^{W\ast}(i,2)g_{L}^{W}(i,3)-B(x_i)
  \frac{m_W}{m_b}g_{L}^{W\ast}(i,2)g_{R}^{G}(i,3)\nn\\
  &+&g_{L}^{G\ast}(i,2)[C(x_i)g_{L}^{G}(i,3)-\frac{m_{u_i}}{m_b}D(x_i)g_{R}^{G}(i,3)]\nonumber\\
  &+&\frac{x_i}{y_i}g_{L}^{{h}\ast}(i,2)[C(y_i)g_{L}^{{h}}(i,3)
   -\frac{m_{u_i}}{m_b}D(y_i)g_{R}^{{h}}(i,3)]\},
  \label{C7}
  \end{eqnarray}
 where $x_i=m_{u_i}^2/m_W^2$ and $y_i=m_{u_i}^2/m_{h^+}^2$. For cross check,
 using the loop functions given in
 the appendix and the CKM matrix unitarity condition, one can easily
 obtain the predication $C_7^{SM}(m_W)= A(x_t)+B(x_t)+x_t[C(x_t)+D(x_t)]$
 which is consistent  with that  in  Ref.~\cite{BLOSM}. Furthermore, $C_7$
 receives a large non-decoupling contribution not only from top quark as in the SM but also
 from the up-type vector-like quark loops at the electroweak scale.
 The non-decoupling effects are unique and will be  demonstrated in next Section.

The Wilson coefficient $C_9$ at the matching scale is
  \begin{eqnarray}
  C_{9}&=&\frac{P(x_t)-Q(x_t)}{\sin^{2}\theta_{W}}
  +4Q(x_t)\nonumber\\
  &-&\frac{2\pi}{\alpha_{em}}\frac{U_{d}^{*52}U_{d}^{53}}{V_{tb}V_{ts}^{\ast}}
  (\frac{1}{4}-\sin^{2}\theta_{W})\nonumber\\
  &+&\frac{1}{V_{tb}V_{ts}^{\ast}}\left\{\sum_{i=3}^{5}\left[R(x_i)g_{L}^{W\ast}(i,2)g_{L}^{W}(i,3)
  +S(x_i)g_{R}^{G\ast}(i,2)g_{L}^{G}(i,3)\right]\right.\nonumber\\
  &+&\sum_{i=1}^{5}\frac{m_W}{m_{u_i}}T(x_i)\left[g_{L}^{W\ast}(i,2)
   g_{L}^{G}(i,3) +g_{R}^{G\ast}(i,2)g_{L}^{W}(i,3)\right]\nn\\
  &+&\left.\frac{x_i}{y_i}S(y_i)g_{R}^{{h}\ast}(i,2)g_{L}^{{h}}(i,3)]\right\}+\frac{4}{9}.
  \label{C9}
  \end{eqnarray}
 Note the first part related to  $P(x_t)$ and $Q(x_t)$
 from the box diagrams and the effective vertex $b\to sZ^*$ at loop
 level have the same
 expression as those in the SM, while  the second part denotes   the interaction at tree level enhanced by
 a large factor $\frac{2\pi}{\alpha_{em}}$. The last part comes from the effective vertex
  $b\to s\gamma^*$ at loop level  for consistency. The contribution from one-loop
 matrix element of the  operator $O_2$ is also included as in the SM \cite{BLOSM}.
Moreover, the Wilson coefficients $C_{10}$, $C'_9$, and $C'_{10}$ at the matching scale are
  \begin{eqnarray}
  C_{10}&=&-\frac{P(x_t)-Q(x_t)}{\sin^{2}\theta_{W}}
  +\frac{2\pi}{\alpha_{em}}\frac{1}{4}\frac{U_{d}^{*52}U_{d}^{53}}{V_{tb}V_{ts}^{\ast}}~,~
  \label{C10}
  \end{eqnarray}
  \begin{eqnarray}
  C_{9}^{'}&=&(\frac{1}{4}-\sin^{2}\theta_{W})\frac{2\pi}{\alpha_{em}}
\frac{V_{d}^{*52}V_{d}^{53}}{V_{tb}V_{ts}^{\ast}}~,~
  \label{C9p}
  \end{eqnarray}
  \begin{eqnarray}
  C_{10}^{'}&=&
  -\frac{2\pi}{\alpha_{em}}\frac{1}{4}\frac{V_{d}^{*52}V_{d}^{53}}{V_{tb}V_{ts}^{\ast}}~.
  \label{C10p}
  \end{eqnarray}
The contributions from loop diagrams to $C'_{9,10}$ can be neglected safely.
\end{itemize}

\begin{itemize}
\item Branching Ratios\\
Considering that the Wilson coefficients do not separate  into the SM and new physics parts
 easily and new operators are introduced, we need to list some explicit expressions for
the branching ratios  of B decays as follows
\begin{enumerate}
\item $B \to X_s\gamma$\\
     The inclusive $B\to  X_s\gamma$  rate is the most precise and
     clean short-distance information that we have, at present,
     on $\Delta B=1$ FCNCs. The new contributions mainly come from the new type Yukawa interactions to operator $O_7$.
     The calculation of the branching ratio is usually normalized by the process
     $B\rightarrow X_{c}e\overline{\nu_{e}}$, so we get
       \begin{equation}
        {\rm Br}(B\rightarrow X_{s}\gamma)={\rm Br}^{ex} (B\rightarrow X_{c}e\overline{\nu_{e}})
        \frac{|V_{ts}^{\ast}V_{tb}|^{2}}{|V_{cb}|^{2}}\frac{6\alpha}
        {\pi f(z)}|C^{eff}_{7}(\mu_b)|^{2}.\label{bsg}
       \end{equation}
     Here $z=\frac{m_c}{m_b}$, and $f(z)=1-8z^2+8z^6-z^8-24z^4\ln z$ is the phase-space factor
     in the semi-leptonic B-decay. From the formula of $C^{eff}_7$ in Eq.(\ref{C7}) and
     the corresponding  coefficients in Eqs. (\ref{ud_1})-(\ref{ud_4}), we can
     see that if we sum  the flavor indices from 1 to 5 in Eqs.~(\ref{ud_1})-(\ref{ud_4}), $C_7$ will be exactly the
     same as the five generation 2HDM. In our numerical calculation we will compare
     both results in these two  models, since it will show clearly the implications of
     the new type Yukawa interactions in the \fsu5 model.
\item $B\to X_s\ell^+\ell^-$\\
     Since the new operators $O_9'$ and $O_{10}'$ contribute
     to $B\to X_s\ell^+\ell^-$ and the exclusive decays,
     the analytical expression of invariant dileptonic mass
     distribution is found to be similar to the SM as follows
    \begin{eqnarray}\label{dgbtt}
    \frac{d\Gamma(B\to X_s\ell^+\ell^-)}{ds}
    &=& \frac{G_F^2m_b^5}{768\pi^5}\alpha_{em}^2\vert V_{tb}V_{ts}^*\vert^2
    (1-s)^2(1-\frac{4r}{s})^{1/2}\nonumber\\
    &\times&  \left\{4|C_7^{eff}|^2(1+\frac{2}{s})+
    (|C_9^{eff}|^2+|C_9'|^2)(1+2s)\right.\nn\\
    &+& \left.(|C_{10}|^2+|C'_{10}|^2)
    (1+2s)+12Re(C_7^{eff}C_9^{eff*})\right\}~,
    \label{bll}
    \end{eqnarray}
    where  $s=(p_{\ell^+}+p_{\ell^-})^2/m_b^2$.
    Also, we use the normalization process $B\rightarrow X_{c}e\overline{\nu_{e}}$
    to get rid of large uncertainties due to $m_b^5$
    and CKM elements as in Eq. (\ref{bsg}).
\item $B_s\to \mu^+\mu^-$\\
     The purely leptonic decays constitute a special case
     among exclusive transitions. It is strongly helicity suppressed and only
     receives contributions from two axial-current operators $O_{10}$ and $O'_{10}$
     in the models we studied. The decay width is given by
     \begin{eqnarray}
     \Gamma (B_s\to \mu^+\mu^-)=\kappa\frac{\alpha_{em}^2G_F^2}
     {16\pi^3}\left\vert V_{tb}V_{ts}^*\right\vert^2f_{B_s}^2m_{B_s}m_\mu^2|C_{10}-C'_{10}|^2~,~\,
     \end{eqnarray}
     where $f_{B_s}$ is the decay constant for $B_s$ determined by
     $\langle 0|\overline{q}\gamma_\mu\gamma_5 b|B_q\rangle=-if_{B_q}p_\mu.$
     The factor $\kappa$ denotes  the
     non-zero width difference of the $B_s$-meson system effect on the branching
     ratio of the $B_s\to \mu^+\mu^-$ decay  and it reads \cite{deBruyn:2012wk}
     \begin{eqnarray}
     \kappa=\frac{1+\frac{1}{2}\tau_{B_s}{\cal A}_{\Delta\Gamma}\Delta\Gamma_s}
     {1-\frac{1}{4}\tau^2_{B_s}(\Delta\Gamma_s)^2},
     \label{kappa}
     \end{eqnarray}
     where $\Delta\Gamma_s$ is the difference between the decay widths of the light
     and heavy $B_s$ mass eigenstates and  $\tau_{B_s}$ is the $B_s$ mean lifetime.
     The parameters ${\cal A}_{\Delta\Gamma}$ is related to the effective $B_s\to \mu^+\mu^-$
     lifetime $\tau_{\mu^+\mu^-}$ and depends sensitively on new physics.
\item $B_s\to \ell^+\ell^-\gamma$\\
     The  exclusive decay can be obtained from the inclusive decay $b\to s\ell^+\ell^-\gamma$,
     and further, from $b\to s\ell^+\ell^-$. To achieve this, for $\ell=e, \mu$ we just attach
     photons to any external quark lines in the Feynman diagrams of $b\to s\ell^+\ell^-$ \cite{xiong08}.
     The decay rate is
     \begin{eqnarray}
     \frac{d\Gamma}{ds}&=&\left|\frac{\alpha_{em}^{3/2}G_F}{4\sqrt{6\pi}}V_{tb}V_{ts}^*\right|^2
     \frac{m^7_{B_s}}{(2\pi)^3}s(1-s)^3\left[|K|^2+|L|^2+|M|^2+|N|^2\right]~,~\,
     \label{gbttg}
     \end{eqnarray}
     where $s=p^2/m_{B_s}^2$ is normalized dileptonic mass squared, and
     \begin{eqnarray}
     K&=&\frac{1}{m_{B_s}^2}\left\{[C_9^{eff}(\mu_b)+C_9']G_1(p^2)-2C_7^{eff}(\mu_b)\frac{m_b}{p^2}G_2(p^2)
     \right\},\nonumber\\
     L&=&\frac{1}{m_{B_s}^2}\left\{[C_9^{eff}(\mu_b)-C_9']F_1(p^2)-2C_7^{eff}(\mu_b)\frac{m_b}{p^2}F_2(p^2)
     \right]~,~\,\nonumber\\
     M&=&\frac{C_{10}+C'_{10}}{m_{B_s}^2}G_1(p^2),\ \ \ \
     N=\frac{C_{10}-C'_{10}}{m_{B_s}^2}F_1(p^2)~,~\,
     \end{eqnarray}
     with  $G_i$ and $\ F_i$ being the form factors~\cite{Eilam95}.
     \end{enumerate}
\end{itemize}

\section{Numerical Results }\label{num}
Since additional vector like quark introduced in the model, there are many
new input parameters appear  in Wilson coefficients  $C_7,\ C_9,\ C_{10},\ C_9',\ C_{10}'$.
These parameters are not independent and constrained by conditions  Eq.~(\ref{massdiag}).
As the first study on B physics in the model,
we will not scan the parameter space completely, but focus on the implication of
mass scale of the vector-like quark on B physics, this will give us the most
important information of the model. Thus in the numerical study we
scan the mass $m_{u_x}$ in the range $180\ {\rm GeV} \sim 2000\ {\rm GeV}$,
and $m_{u_y}$  in the range $40\sim 60$ GeV heavier than $m_{u_x}$. As for other parameters,
we use the shooting method to randomly generate $5\times 5$ unitary
matrix $V_u$ and $U_u$, then use the CKM matrix to get the $V_d$,  $U_d$ to let
mass of down-type quark matrix satify the Eq.~(\ref{massdiag}).
Note that to take in account impact of the non-zero width difference of
$B_s$ system \cite{Aaij:2012kn} on the branching ratio of $B_s\to \mu^+\mu^-$,
we use $y_s= 0.088\pm 0.014$ \cite{deBruyn:2012wk}.
We also use the following experimental constraints from B physics:
\begin{enumerate}
\item In the model with three generation quarks, the CKM matrix unitarity
     is already used in the calculations of the loop-level FCNC induced rare B decays.
     Therefore for consistency, in the model we study
     the constraints on CKM matrix element measurements  are not from rare B decays but
     from tree-level B decays~\cite{CKM} as shown in Table~\ref{tab:c1B}.
     \begin{table}[htb]
     \caption[]{The CKM matrix elements constrained by the tree-level B decays.}
     \label{tab:c1B}
     \begin{center}
       \begin{tabular}{|c||c|c|c|}
        \hline
        & absolute value  & relative error & direct measurement from \\ \hline
        $V_{ud}$ & $0.97418 \pm 0.00027$ & $0.028\%$& nuclear beta decay \\\hline
        $V_{us}$ & $0.2255  \pm 0.0019$  & $0.84\%$ & semi-leptonic K-decay\\\hline
        $V_{ub}$ & $0.00393 \pm 0.00036$ & $9.2\%$  & semi-leptonic B-decay\\\hline
        $V_{cd}$ & $0.230   \pm 0.011$   & $4.8\%$  & semi-leptonic D-decay\\\hline
        $V_{cb}$ & $0.0412  \pm 0.0011$  & $2.7\%$  & semi-leptonic B-decay\\\hline
        $V_{tb}$ & $>0.74$               &          & (single) top-production\\\hline
       \end{tabular}
     \end{center}
     \end{table}
     \item To see the implications of the vector-like quark multiplets,
     we use the following bounds on the rare B decays~\cite{Bmeasured,Aaij:2012kn}
        \begin{eqnarray}
     &&   Br(b\to ce\overline{\nu}_{e})= (10.74 \pm 0.16) \times 10^{-2}~,~\nn\\
     &&   Br({\overline B}\to X_s \gamma)= (3.06 \pm 0.23) \times 10^{-4}~,~\nn\\
     &&   Br(B\to X_s\ell^+\ell^-)= (4.5 \pm 1) \times 10^{-6}~,~\nn\\
     &&   Br( B_s\to \mu^+\mu^-)< 4.5 \times 10^{-9}\ \   (95\% C.L.)~.
     \end{eqnarray}
\item Other input parameters are the same as those in the SM, except for $\tan\beta$
     and  the charged Higgs boson mass $m_{h^+}$. In our numerical
     calculations we scan the two parameters randomly and choose
     two typical points ($\tan\beta=2,~~ m_{h^+}={3000\ {\rm GeV}}$) and
     ($\tan\beta=40,~~ m_{h^+}={500\ {\rm GeV}}$) for the demonstration.
\end{enumerate}

\begin{figure}[hbtp]
\epsfig{file=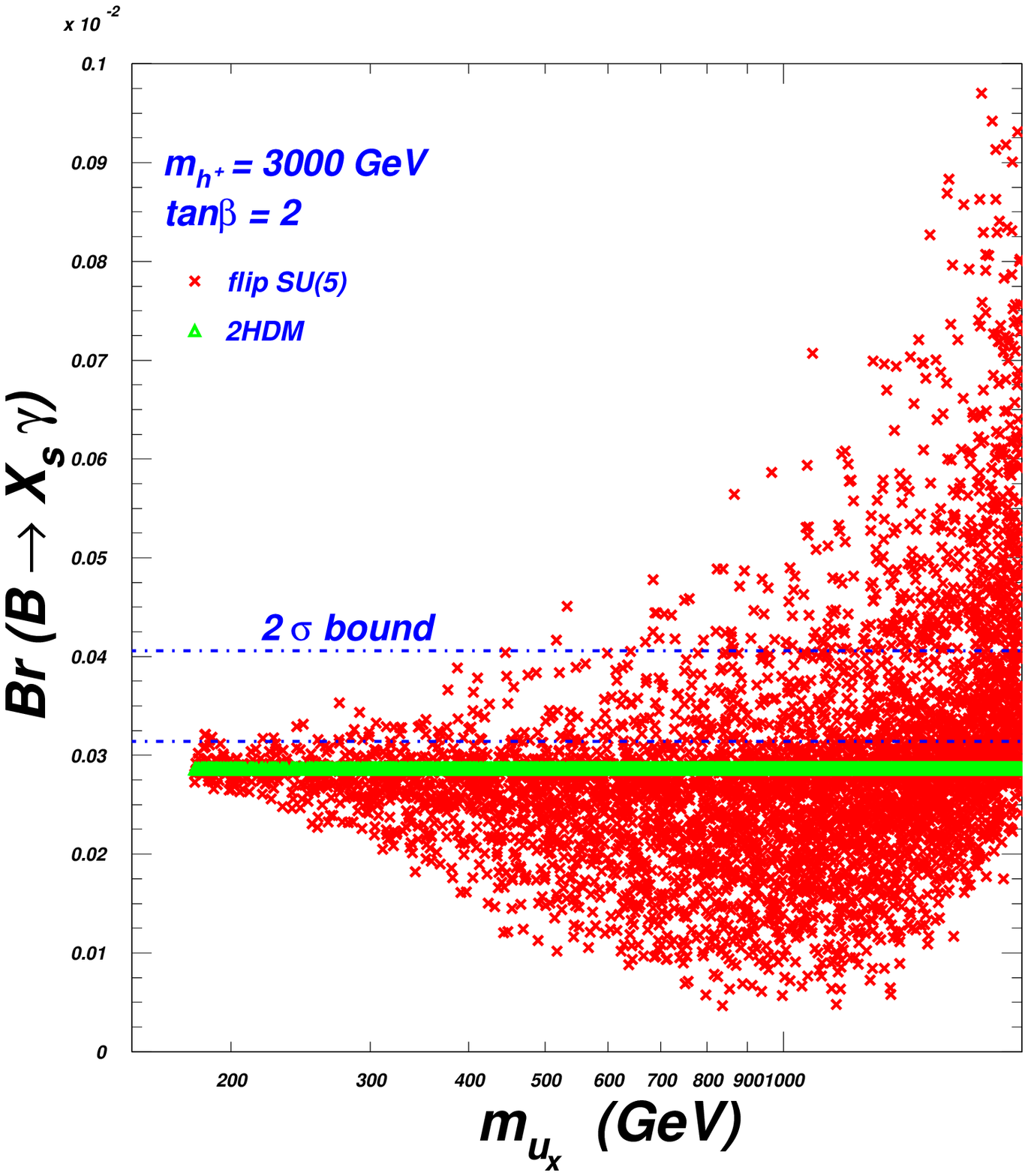,height=8cm}
\hspace{-1.45cm} \epsfig{file=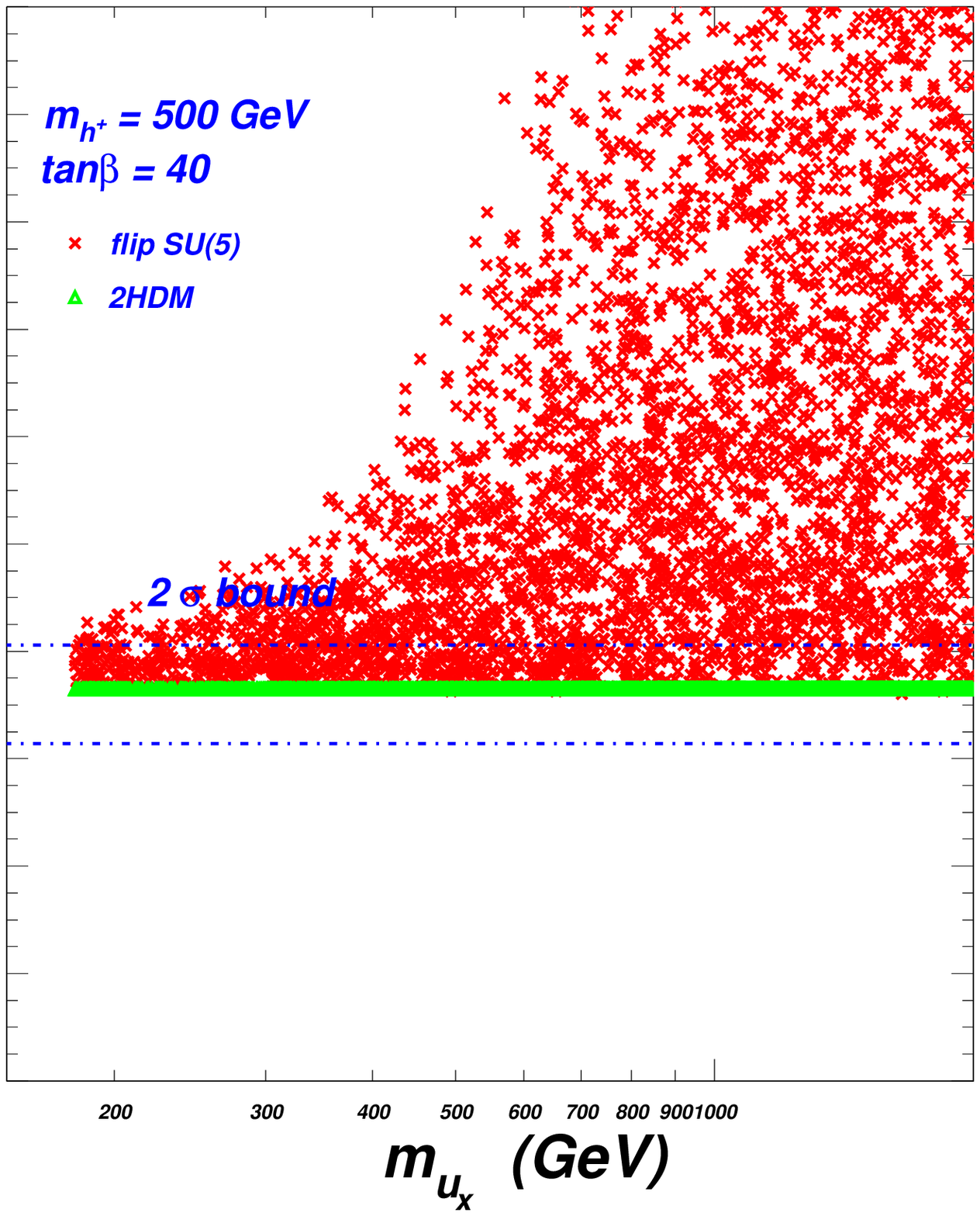,height=8cm}
\vspace{-0.3cm}
\caption{Comparison of $B\to X_s \gamma$ versus $m_{u_x}$
in the \fsu5 model (red cross) and 2HDM (green triangle).}
   \label{fig1}
\end{figure}

The numerical results of $B\to X_s \gamma$ as a function of the vector-like quark mass are displayed
in Fig.~\ref{fig1}. For the comparison, Fig.~\ref{fig1}  also shows the results
of the five-generation 2HDM. From this figure one can see some features clearly:
(i) The new physics effects decouple when the charged Higgs boson is very heavy. However,
for a much heavier charged Higgs,
the branching ratio of $B\to X_s \gamma$ increases with
 $m_{u_x}$  in the \fsu5 model
while is almost independent on the extra quark  mass in 2HDM, indicating the large non-decoupling effects;
(ii) Unlike the 2HDM where the large $\tan\beta$ is  preferred if the charged Higgs boson
mass is at the EW scale,  the small $\tan\beta$, which is excluded
in 2HDM, is still survived  in the \fsu5 model; (iii)
It is clear from the left plot of this figure that the branching ratio can be much bigger than the detection
result when $m_{u_x}$ getting close to the charged
 Higgs boson mass. So the detection results of  $B\to X_s \gamma$
can give stringent constraints on the \fsu5 model.
The tendency of the figure can be understood as following:

\begin{itemize}
\item  $C_7$ determined by  Eq.~(\ref{C7}) in both \fsu5 model and 2HDM \cite{BNP2HDM}
      will approach to the SM value when the charged Higgs boson is much heavier than
     EW scale. Nevertheless, the contributions from the fourth and fifth
     generation up-type vector-like quarks in 2HDM  can be suppressed by small $V^{5i}$ and $V^{4i}$
     due to the unitarity condition of  $5\times 5$ matrix;
\item Because the summed indices are only from 1 to 4 in the \fsu5 model,
     the unitary condition of the CKM matrix can not be maintained. When
     the vector-like particle mass  approaches to the charged Higgs boson mass,
      the suppression from
     $5\times 5$ CKM mixing matrix will be released and then the non-decoupling effects will be sizable.
     In fact, the non-decoupling effects are a very
     special part of the \fsu5 model at EW scale and can be tested
     at the LHC and other B physics detectors.
\end{itemize}

 \begin{figure}[htbp]
 \epsfig{file=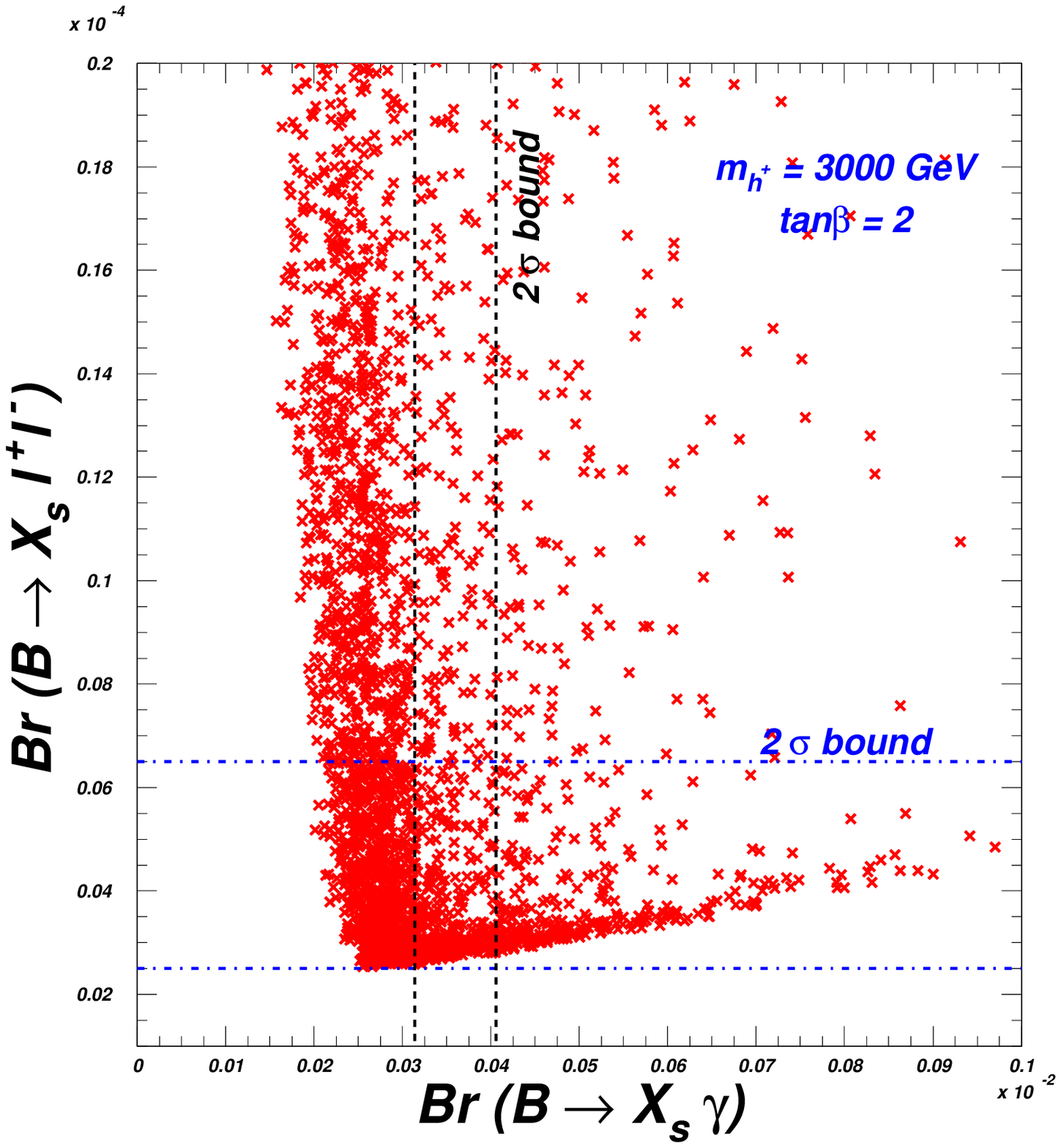,height=8cm}
 \epsfig{file=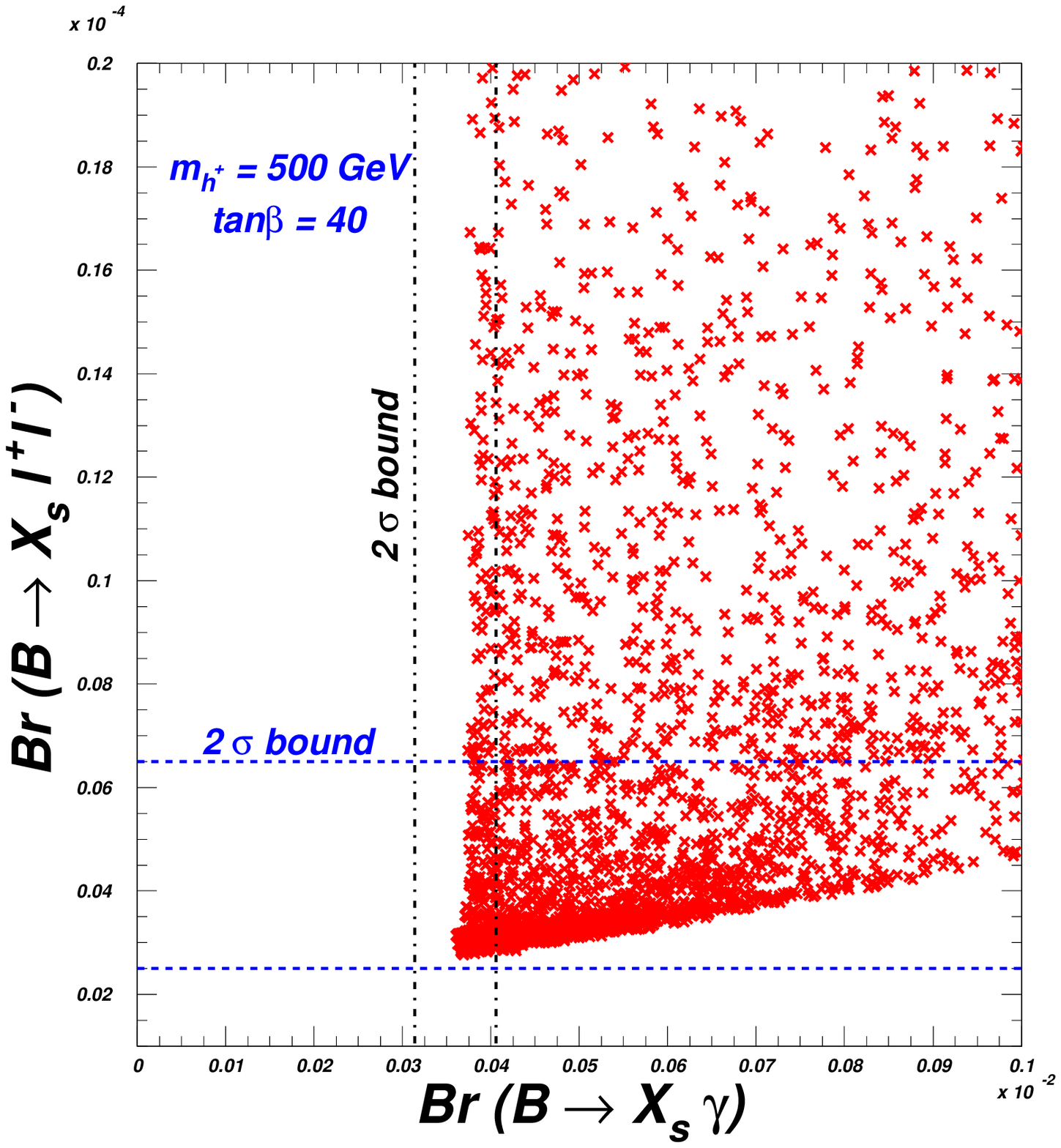,height=8cm}
 \vspace{-0.3cm}
 \caption{Branching ratio of $B \to X_s \ell^+\ell^-$
 versus $B\to X_s \gamma $ in the \fsu5 model.}
   \label{fig2}
\end{figure}
\begin{figure}[htbp]
 \epsfig{file=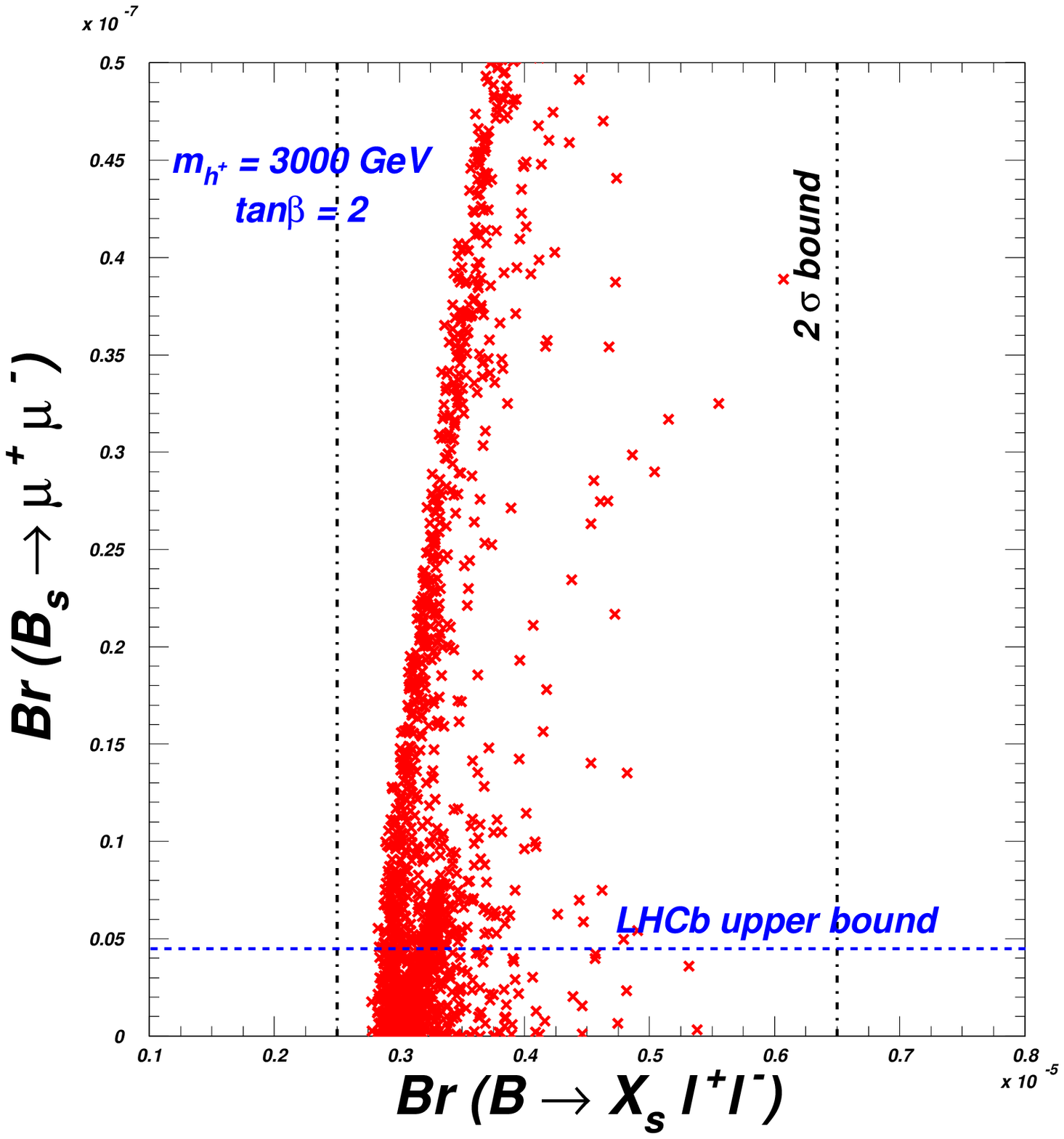,height=8cm}
 \epsfig{file=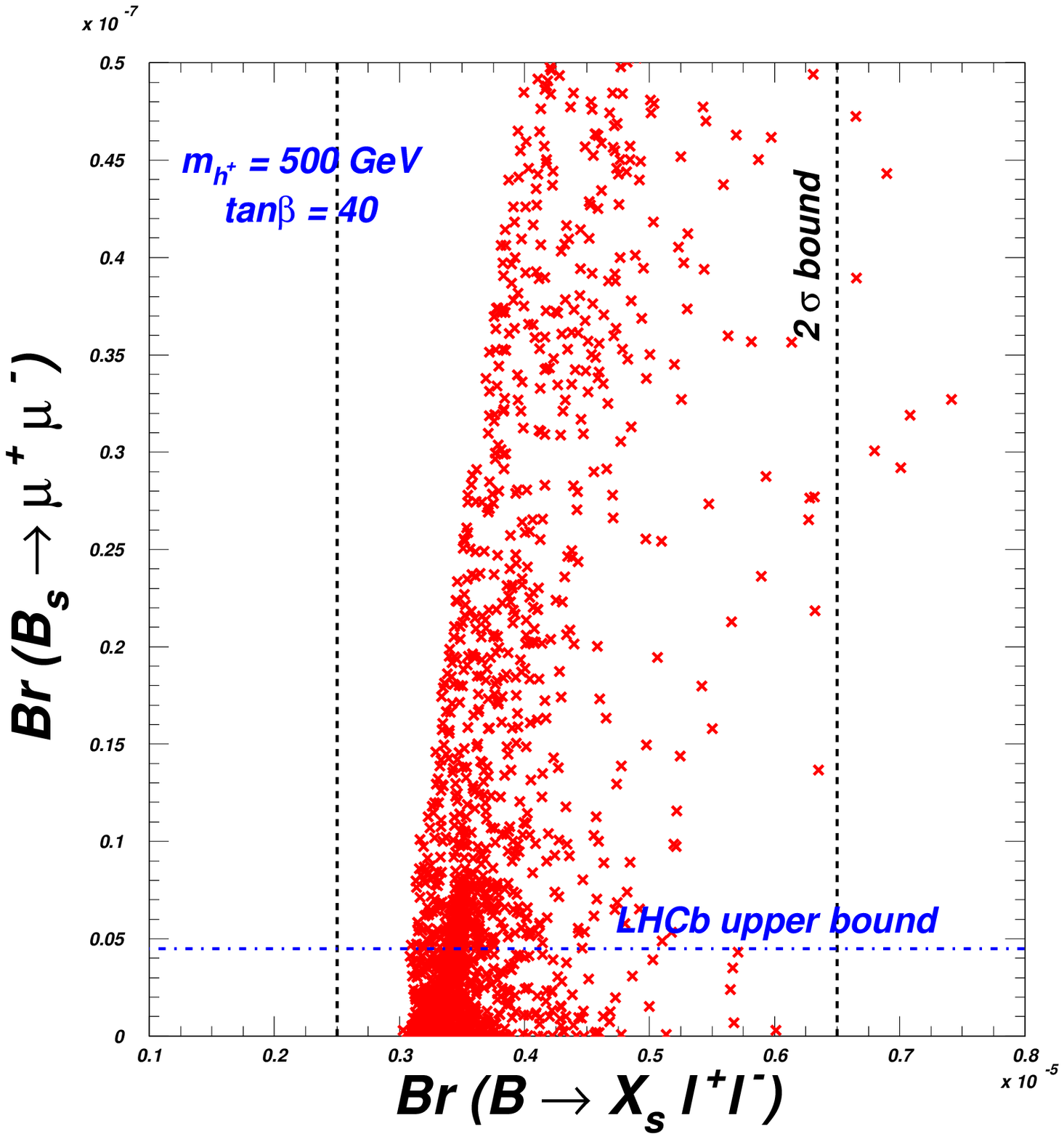,height=8cm}
 \vspace{-0.3cm}
 \caption{Branching ratios of $B_s \to \mu^+ \mu^- $ versus
 $B \to X_s \ell^+\ell^-$ in the \fsu5 model.}
   \label{fig3}
\end{figure}

Fig.~\ref{fig2} shows the branching ratio of $B \to X_s \ell^+\ell^-$
versus $B\to X_s \gamma $ in the \fsu5 model.  Clearly, both processes will
give stringent constraints
on our model. Especially, most part of the points are excluded when the charged
Higgs boson is several hundred GeV,
leaving a narrow part in the parameter space.
Similar phenomenology can be seen in Fig.~\ref{fig3}
which shows branching ratios of $B_s \to \mu^+ \mu^- $ versus
$B \to X_s \ell^+\ell^-$.  The non-decoupling effects can
be stringently constrained  by the experiments as expected.
 Here we should emphasize that the upper bounds from
the Tevatron and the first LHCb constraints~\cite{Aaij:2012kn},
which are about one order of magnitude above the SM expectation, as well as the recent
CDF results of $B_s \to \mu^+ \mu^- $ detection~\cite{Aaltonen:2011fi}
 can be  explained naturally.
It is interesting to see that there is an approximate linear
relation between branching ratios of $B_s \to \mu^+ \mu^- $ and
$B \to X_s \ell^+\ell^-$. In fact, we find that in the allowed parameter space with
$U_d\simeq V_d^\dagger$, the dominant contributions to both processes
come from $C_{i}$ and $C_{i}^{'}(i=9,10)$. From Eqs. (\ref{C9}) to (\ref{C10p}), we
can easily draw the conclusion that the branching
ratios are nearly proportional to $|C_{10}^{'}|^2$.

\begin{figure}[htbp]
 \epsfig{file=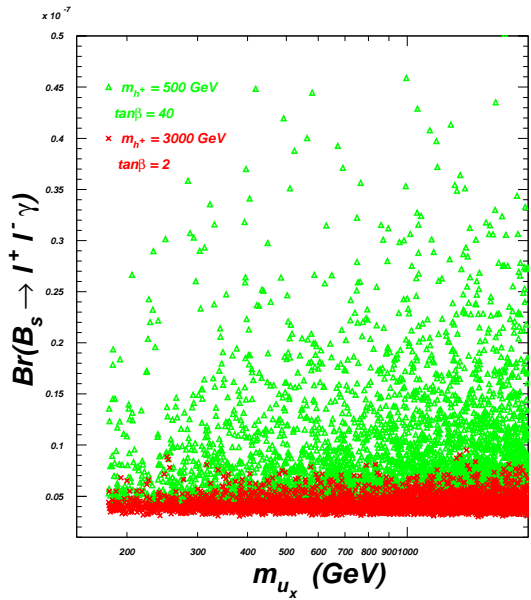,height=8cm}
 \vspace{-0.3cm}
 \caption{Branching ratio of $B_s \to \ell^+ \ell^-\gamma$ with the combained constraints from
$B\to X_s \gamma$, $B \to X_s \ell^+\ell^-$ and $B_s \to \mu^+ \mu^- $.
Red cross stands for the type inputs ($\tan\beta=2,~~ m_{h^+}={3000\rm GeV}$) and
 green triangle for ($\tan\beta=40,~~ m_{h^+}={500\rm GeV}$) in the \fsu5 model, respectively.}
   \label{fig4}
\end{figure}

To see whether there are solutions
simultaneously satisfied with the allowed ranges for these data, we can offer now some predications
 for $B_s \to \ell^+\ell^- \gamma $, which might  be
measured at the LHCb and B factories.  The numerical results are illustrated  in Fig.~\ref{fig4}.
We can see clearly that under the constraints from the inclusive decays
$B\to X_s \gamma$ and $B \to X_s \ell^+\ell^-$,
exclusive decays $B_s \to \mu^+ \mu^- $, as well as CKM measurements
 extracted by the tree-level B decays, the branching ratio,
which is very sensitive to $\tan \beta$ and charged Higgs
boson mass, can still be up to $(4\sim5)\times10^{-8}$.
Thus, it may be tested by the LHCb soon.

Rare B decays continue to be the valuable probes of physics beyond the SM.
In the current early phase of the LHC era, the exclusive modes with muons
in the final states are among the most promising decays. The decay $B_s\to \mu^+\mu^-$
is likely to be confirmed before the end of 2012~\cite{LHCb}.
If an enhancement beyond $10^{-8}$ and further non-decoupling effects are  observed,
we will have an indication of the \fsu5 model.
Although there are some theoretical challenges including calculation of  the
hadronic form factors and non-factorable corrections,
$B_s\to \ell^+\ell^-\gamma$  can be expected as the next goal once $B_s\to \mu^+\mu^-$
measurement is finished since the final states can be identified easily and branching ratios are large.
Our predictions for such processes can be tested in the near future.

\section{Summary}\label{sum}
In this paper, we studied the vector-like quark contributions to B physics processes
in the \fsu5 model, including the quark mass spectra,
Feynman rules, the new operators in low energy effective theory and the correspondence
Wilson coefficients, etc. As for the first time study, we focus on the implication of
mass scale of vector like quark.  The main conclusions we obtained are the following:
\begin{enumerate}
\item There exists the  $\overline{s}bZ$ interaction  at tree level,
      and the Yukawa interactions are changed.
      The new operators $O_9'$ and $O_{10}'$ must be introduced in
      effective Hamiltonian, and the Wilson coefficients  are
      changed due to the violation of the unitarity condition.
\item Different from many previous models, the effects of vector-like
      quarks on rare B decays such as $B\to X_s\gamma$ and $B\to X_s\ell^+\ell^-$
      do not decouple in some allowed parameter space, especially
      when the vector-like quark mass is comparable to the charged Higgs
      boson mass.
\item Under the constraints  from $B\to X_s\gamma$ and $B\to X_s\ell^+\ell^-$,
      there exist scenarios in the model the latest measurement for
      $B_s\to \mu^+\mu^-$ can be explained naturally,
      and the branching ratio of $B_s\to \ell^+\ell^-\gamma$ can be up
      to $(4\sim5)\times10^{-8}$.
\end{enumerate}

All in all, due to the participation of  vector-like particles, the \fsu5 model is
different from the ordinary models such as 2HDM. In particular, the non-decouling effects
are much more predictable and may be tested in the near future experiments.
Finally, we should note that the large input parameter space and the
sparticle effects in the \fsu5 model needs further work.
\begin{acknowledgments}
This research was supported in part by the Natural Science Foundation of China
under grant numbers 11005006, 11172008, 10821504, 11075194, and 11135003,
 by the DOE grant DE-FG03-95-Er-40917,
 and by the Doctor
Foundation of BJUT No. X0006015201102.
\end{acknowledgments}

\section*{Appendix}
The loop functions for calculating the Wilson coefficients at the matching scale are the following
\begin{eqnarray}
A(x)&=&\frac{5x+38x^2-55x^{2}}{36(x-1)^{3}}+\frac{4x-17x^{2}+15x^{3}}{6(x-1)^{4}}\ln x,\nn\\
B(x)&=&\frac{x+x^2}{4(x-1)^{2}}-\frac{x^{2}}{2(x-1)^{3}}\ln x,\nn\\
C(x)&=&\frac{20-19x+5x^{2}}{18(x-1)^3}+\frac{-2+x}{3(x-1)^{4}}\ln x\nn\\
D(x)&=&\frac{-5-5x+4x^2}{12(x-1)^{3}}+\frac{2x-x^2}{2(x-1)^{4}}\ln x,\nn\\
P(x)&=&\frac{-x}{4(x-1)}+\frac{x}{4(x-1)^{2}}\ln x~,\nn\\
Q(x)&=&\frac{x^2-6x}{8(x-1)}+\frac{3x^2+6x}{8(x-1)^{2}}\ln x~,\nn\\
R(x)&=&\frac{31x^{2}+20x^{3}}{9(x-1)^{3}}+\frac{-4+18x-30x^{2}+6x^{3}}{9(x-1)^{4}}\ln x,\nn\\
S(x)&=&\frac{38-79x+47x^{2}}{108(x-1)^{3}}+\frac{-4x+6x^{2}-3x^{4}}{18(x-1)^{4}}\ln x,\nn\\
T(x)&=&\frac{x-5x^{2}-2x^{3}}{12(x-1)^{3}}+\frac{x^{3}}{2(x-1)^{4}}\ln  x.
\end{eqnarray}


\begin{thebibliography}{99}
\bibitem{Langacker:1991an}
  J.~R.~Ellis, S.~Kelley and D.~V.~Nanopoulos,
  Phys.\ Lett.\ B {\bf 260}, 131 (1991);
  P.~Langacker and M.~X.~Luo,
  Phys.\ Rev.\ D {\bf 44}, 817 (1991);
  U.~Amaldi, W.~de Boer and H.~Furstenau,
  Phys.\ Lett.\ B {\bf 260}, 447 (1991);
 F.~Anselmo, L.~Cifarelli, A.~Peterman and A.~Zichichi,
  Nuovo Cim.\  A {\bf 104}, 1817 (1991);
  Nuovo Cim.\  A {\bf 105}, 1025 (1992).


\bibitem{Ellis:1983ew}
  J.~R.~Ellis, J.~S.~Hagelin, D.~V.~Nanopoulos, K.~A.~Olive, M.~Srednicki,
  Nucl.\ Phys.\  {\bf B238}, 453-476 (1984).

\bibitem{Goldberg:1983nd}
  H.~Goldberg,
  Phys.\ Rev.\ Lett.\  {\bf 50}, 1419 (1983).

\bibitem{smbarr} S. M. Barr,
Phys.\ Lett.\ B {\bf 112}, 219 (1982).


\bibitem{dimitri}
J.~P.~Derendinger, J.~E.~Kim and D.~V.~Nanopoulos,
Phys.\ Lett.\ B {\bf 139}, 170 (1984).

\bibitem{AEHN-0}
  I.~Antoniadis, J.~R.~Ellis, J.~S.~Hagelin and D.~V.~Nanopoulos,
  Phys.\ Lett.\  B {\bf 194}, 231 (1987).


\bibitem{Jiang:2006hf}
  J.~Jiang, T.~Li and D.~V.~Nanopoulos,
  Nucl.\ Phys.\  B {\bf 772}, 49 (2007).


\bibitem{Antoniadis:1988tt}
  I.~Antoniadis, J.~R.~Ellis, J.~S.~Hagelin and D.~V.~Nanopoulos,
  Phys.\ Lett.\  B {\bf 208}, 209 (1988)
  [Addendum-ibid.\  B {\bf 213}, 562 (1988)];
  Phys.\ Lett.\  B {\bf 231}, 65 (1989).


\bibitem{Lopez:1992kg}
  J.~L.~Lopez, D.~V.~Nanopoulos and K.~J.~Yuan,
  Nucl.\ Phys.\  B {\bf 399}, 654 (1993).


\bibitem{Beasley:2008dc}
  C.~Beasley, J.~J.~Heckman and C.~Vafa,
  JHEP {\bf 0901}, 058 (2009);
  JHEP {\bf 0901}, 059 (2009);
R.~Donagi and M.~Wijnholt,
  arXiv:0802.2969 [hep-th];
  arXiv:0808.2223 [hep-th].

\bibitem{Jiang:2009zza}
  J.~Jiang, T.~Li, D.~V.~Nanopoulos and D.~Xie,
  Phys.\ Lett.\  B {\bf 677}, 322 (2009);
  Nucl.\ Phys.\  B {\bf 830}, 195 (2010).

\bibitem{Nakamura:2003hk}
  K.~Nakamura,
  Int.\ J.\ Mod.\ Phys.\  A {\bf 18}, 4053 (2003).

\bibitem{DUSEL}
  S.~Raby {\it et al.},
  arXiv:0810.4551 [hep-ph].

\bibitem{Li:2009fq}
  T.~Li, D.~V.~Nanopoulos and J.~W.~Walker,
  Phys.\ Lett.\  B {\bf 693}, 580 (2010).

\bibitem{Li:2010dp}
  T.~Li, D.~V.~Nanopoulos and J.~W.~Walker,
  Nucl.\ Phys.\  B {\bf 846}, 43 (2011)
  [arXiv:1003.2570 [hep-ph]].

\bibitem{Kyae:2005nv}
  B.~Kyae and Q.~Shafi,
  Phys.\ Lett.\  B {\bf 635}, 247 (2006)
  [arXiv:hep-ph/0510105].

\bibitem{Huo:2011zt}
  Y.~Huo, T.~Li, D.~V.~Nanopoulos and C.~Tong,
  arXiv:1109.2329 [hep-ph].

\bibitem{Li:2011ab}
  T.~Li, J.~A.~Maxin, D.~V.~Nanopoulos and J.~W.~Walker,
  Phys.\ Lett.\ B {\bf 710}, 207 (2012)
  [arXiv:1112.3024 [hep-ph]].

\bibitem{Cremmer:1983bf}
  E.~Cremmer, S.~Ferrara, C.~Kounnas and D.~V.~Nanopoulos,
  Phys.\ Lett.\  B {\bf 133}, 61 (1983);
J.~R.~Ellis, A.~B.~Lahanas, D.~V.~Nanopoulos and K.~Tamvakis,
  Phys.\ Lett.\  B {\bf 134}, 429 (1984);
J.~R.~Ellis, C.~Kounnas and D.~V.~Nanopoulos,
  Nucl.\ Phys.\  B {\bf 241}, 406 (1984);
  Nucl.\ Phys.\  B {\bf 247}, 373 (1984);
A.~B.~Lahanas and D.~V.~Nanopoulos,
  Phys.\ Rept.\  {\bf 145}, 1 (1987).

\bibitem{Li:2010ws}
  T.~Li, J.~A.~Maxin, D.~V.~Nanopoulos and J.~W.~Walker,
  Phys.\ Rev.\  D {\bf 83}, 056015 (2011)
  [arXiv:1007.5100 [hep-ph]].


\bibitem{Li:2010mi}
  T.~Li, J.~A.~Maxin, D.~V.~Nanopoulos and J.~W.~Walker,
  Phys.\ Lett.\  B {\bf 699}, 164 (2011)
  [arXiv:1009.2981 [hep-ph]].

\bibitem{Li:2012uj}
  T.~Li, J.~A.~Maxin, D.~V.~Nanopoulos and J.~W.~Walker,
  arXiv:1202.0509 [hep-ph].

\bibitem{BNPfourth}
  A.~Soni, A.~K.~Alok, A.~Giri, R.~Mohanta and S.~Nandi,
  Phys.\ Rev.\ D {\bf 82}, 033009 (2010);
  A.~J.~Buras, B.~Duling, T.~Feldmann, T.~Heidsieck, C.~Promberger and S.~Recksiegel,
  JHEP {\bf 1009}, 106 (2010);
  O.~Eberhardt, A.~Lenz and J.~Rohrwild,
  Phys.\ Rev.\ D {\bf 82}, 095006 (2010);
  Z. H.~Xiong,
  High Energy Phys.\ Nucl.\ Phys.\  {\bf 30}, 284 (2006).


\bibitem{BLOSM}
  A.~J.~Buras, M.~Misiak, M.~M\"unz and S.~Pokorski,
  Nucl.\ Phys.\ B {\bf 424}, 374 (1994).


\bibitem{BHOSM}
  M.~Misiak,
{\it et al.},
  Phys.\ Rev.\ Lett.\  {\bf 98}, 022002 (2007).
  T.~Hurth,
  Rev.\ Mod.\ Phys.\  {\bf 75}, 1159 (2003);
  C.~Bobeth, P.~Gambino, M.~Gorbahn and U.~Haisch,
  JHEP {\bf 0404}, 071 (2004);
  A.~Ghinculov, T.~Hurth, G.~Isidori and Y.~P.~Yao,
  Nucl.\ Phys.\ B {\bf 685}, 351 (2004);
  H.~H.~Asatryan, H.~M.~Asatrian, C.~Greub and M.~Walker,
  Phys.\ Rev.\ D {\bf 65}, 074004 (2002).


\bibitem{BNPMSSM}
  P.~H.~Chankowski and \L.~S\l awianowska,
  Phys.\ Rev.\ D {\bf 63}, 054012 (2001);
  P.~L.~Cho, M.~Misiak and D.~Wyler,
  Phys.\ Rev.\ D {\bf 54}, 3329 (1996);
  Y.~Grossman, Z.~Ligeti and E.~Nardi,
  Phys.\ Rev.\ D {\bf 55}, 2768 (1997);
  J.~L.~Hewett and J.~D.~Wells,
  Phys.\ Rev.\ D {\bf 55}, 5549 (1997);
  S.~Bertolini, F.~Borzumati, A.~Masiero and G.~Ridolfi,
  Nucl.\ Phys.\ B {\bf 353}, 591 (1991);
  A.~J.~Buras and M.~M\"unz,
  Phys.\ Rev.\ D {\bf 52}, 186 (1995);
  M.~Ciuchini, G.~Degrassi, P.~Gambino and G.~F.~Giudice,
  Nucl.\ Phys.\ B {\bf 527}, 21 (1998);
  C.~-S.~Huang, W.~Liao and Q.~-S.~Yan,
  Phys.\ Rev.\ D {\bf 59}, 011701 (1999);
  C.~-S.~Huang and S.~-H.~Zhu,
  Phys.\ Rev.\ D {\bf 61}, 015011 (2000);
  C.~-S.~Huang, W.~Liao, Q.~-S.~Yan and S.~-H.~Zhu,
  Phys.\ Rev.\ D {\bf 63}, 114021 (2001);
  Eur.\ Phys.\ J.\ C {\bf 25}, 103 (2002);
  S.~R.~Choudhury and N.~Gaur,
  Phys.\ Lett.\ B {\bf 451}, 86 (1999).


\bibitem{BNP2HDM}
  B.~Grinstein, R.~P.~Springer and M.~B.~Wise,
  Phys.\ Lett.\ B {\bf 202}, 138 (1988);
  B.~Grinstein, R.~P.~Springer and M.~B.~Wise,
  Nucl.\ Phys.\ B {\bf 339}, 269 (1990);
  Y.~-B.~Dai, C.~-S.~Huang and H.~-W.~Huang,
  Phys.\ Lett.\ B {\bf 390}, 257 (1997);
  J.~L.~Hewett,
  Phys.\ Rev.\ D {\bf 53}, 4964 (1996);
  H.~E.~Logan and U.~Nierste,
  Nucl.\ Phys.\ B {\bf 586}, 39 (2000);
  C.~Bobeth, T.~Ewerth, F.~Kruger and J.~Urban,
  Phys.\ Rev.\ D {\bf 64}, 074014 (2001);
  G.~Erkol and G.~Turan,
  Phys.\ Rev.\ D {\bf 65}, 094029 (2002);
  G.~K.~Yeghiyan,
  Mod.\ Phys.\ Lett.\ A {\bf 16}, 2151 (2001);
  A.~Diaz Rodolfo, R.~Martinez and J.~A.~Rodriguez,
  Phys.\ Rev.\ D {\bf 64}, 033004 (2001);
  R.~Diaz, R.~Martinez and J.~A.~Rodriguez,
  Phys.\ Rev.\ D {\bf 63}, 095007 (2001);
  S.~Davidson and H.~E.~Haber,
  Phys.\ Rev.\ D {\bf 72}, 035004 (2005);
  L.~Wolfenstein and Y.~L.~Wu,
  Phys.\ Rev.\ Lett.\  {\bf 73}, 2809 (1994);
  T.~M.~Aliev and M.~Savci,
  Phys.\ Lett.\ B {\bf 452}, 318 (1999).


\bibitem{Bmeasured}
  D.~Asner {\it et al.}  [Heavy Flavor Averaging Group Collaboration]
    and updates at http://www.slac.stanford.edu/xorg/hfag/


\bibitem{Chatrchyan:2012rg}
  S.~Chatrchyan {\it et al.}  [CMS Collaboration],
  arXiv:1203.3976 [hep-ex].

\bibitem{xiong08}
  Z.~Heng, R.~J.~Oakes, W.~Wang, Z.~Xiong and J.~M.~Yang,
  Phys.\ Rev.\ D {\bf 77}, 095012 (2008);
  Z.~Xiong and J.~M.~Yang,
  Nucl.\ Phys.\ B {\bf 628}, 193 (2002);
  G.~Buchalla and A.~J.~Buras,
  Nucl.\ Phys.\ B {\bf 400}, 225 (1993);
  D.~-S.~Du, C.~Liu and D.~-X.~Zhang,
  Phys.\ Lett.\ B {\bf 317}, 179 (1993).

\bibitem{deBruyn:2012wk}
  K.~de Bruyn, R.~Fleischer, R.~Knegjens, P.~Koppenburg, M.~Merk, A.~Pellegrino and N.~Tuning,
  arXiv:1204.1737 [hep-ph].

\bibitem{Eilam95}
  G.~Eilam, I.~E.~Halperin and R.~R.~Mendel,
  Phys.\ Lett.\ B {\bf 361}, 137 (1995).

\bibitem{Aaij:2012kn}
  R.~Aaij {\it et al.}  [LHCb Collaboration],
  Phys.\ Lett.\ B {\bf 707}, 349 (2012)
  [arXiv:1111.0521 [hep-ex]];
  Phys.\ Lett.\ B {\bf 699}, 330 (2011),
  arXiv:1203.4493 [hep-ex].

\bibitem{CKM}
  K. Nakamura et al. (Particle Data Group), J. Phys. G {\bf 37}, 075021 (2010)
  and 2011 partial update for the 2012 edition.

\bibitem{Aaltonen:2011fi}
  T.~Aaltonen {\it et al.}  [CDF Collaboration],
  Phys.\ Rev.\ Lett.\  {\bf 107}, 239903 (2011)
  [Phys.\ Rev.\ Lett.\  {\bf 107}, 191801 (2011)]
  [arXiv:1107.2304 [hep-ex]].
\bibitem{LHCb} Talk by M. Palutan at Beauty 2011, Amsterdam, April 5, (2011).
\end{thebibliography}
\end{document}